\documentclass[aps,prl,showpacs,twocolumn]{revtex4}
\usepackage{amsmath, amssymb}

\newtheorem{lemma}{Lemma}
\newtheorem{corollary}{Corollary}
\newtheorem{theorem}{Theorem}

\begin{document}
\title{Verifying continuous-variable entanglement in finite spaces}

\author{J. Sperling} \email{jan.sperling2@uni-rostock.de}
\author{W. Vogel} \email{werner.vogel@uni-rostock.de}
\affiliation{Arbeitsgruppe Quantenoptik, Institut f\"ur Physik, Universit\"at Rostock, D-18051 Rostock, Germany}


\begin{abstract}
Starting from arbitrary Hilbert spaces, we reduce the problem to verify entanglement of any bipartite quantum state to finite dimensional subspaces.
Hence, entanglement is a finite dimensional property.
A generalization for multipartite quantum states is also given.
\end{abstract}

\date{\today}

\maketitle

\section{Introduction}
Quantum entanglement has been studied already in the early days of quantum physics as a nonclassical correlation between the subsystems of a compound quantum system.
For example, Einstein, Podolsky and Rosen considered entanglement to be the result of the  incompleteness of quantum theory~\cite{ERP}.
The consequences in the macroscopic worlds were addressed by Schr\"odinger in the context of his cat paradox~\cite{cat}.
Nowadays entanglement is considered to be the key resource of rapidly crowing fields 
of research, such as Quantum Information and Quantum Technology, for example see~\cite{book1}.
Entangled states are required, for example, for quantum key distribution, quantum dense coding, and quantum teleportation~\cite{ekert91,bennett-wiesner92,bennett93}.

Entanglement can be identified by applying all positive but not completely positive (PNCP) maps to a given state~\cite{physlettA223-1}.
The presently best studied PNCP map is the partial transposition (PT)~\cite{prl-77-1413}.
This map can be applied to arbitrary Hilbert spaces.
It is known that PT is necessary and sufficient to verify entanglement for $2\otimes 2$ and $2\otimes 3$ dimensional systems~\cite{physlettA223-1}.
For finite dimensional spaces other PNCP maps are known, which may identify entanglement in cases when the PT map fails.
Example of such maps have recently been proposed by Kossakowski, Ha, Breuer, Hall~\cite{Koss,Ha,Breuer,Hall}, and the realignment criterion for finite spaces~\cite{Realign1,Realign2}.
However, the general form of PNCP maps is yet unknown.

A more sophisticated problem is the task of identifying entanglement of quantum states for continuous variables.
Even if the PNCP maps would be known in this case, the application to test general quantum states cannot be implemented numerically in a straightforward manner.
In infinite dimensional spaces the PT map provides a necessary and sufficient entanglement test for bipartite Gaussian quantum states only~\cite{Duan,Simon}.
A general and complete test of PT entanglement in infinite Hilbert spaces has been proposed in terms of observable moments~\cite{Shchukin}.
However, this gives only a sufficient condition for entanglement.
The identification of infinite dimensional bound entangled states is unresolved so far, cf.~\cite{Miran}.

Equivalent to the PNCP approach is the identification of entanglement with Hermitian operators.
An entanglement witness is introduced as a linear operator, whose mean values are non-negative for separable states but it can become negative for entangled states~\cite{physlettA223-1}.
Recently we have introduced, guided by the general structure of the linear entanglement witnesses, a method to construct necessary and sufficient optimized entanglement conditions for quantum states in an arbitrary Hilbert space in terms of general Hermitian operators~\cite{vospe}.

In the present contribution we show, that for any entangled state there exist finite subspaces of each individual system, where the statistical operator is entangled as well.
Thus entanglement tests of infinite dimensional systems can be reduced in general to finite dimensional ones. Applying the known optimized entanglement conditions~\cite{vospe}, such test can be readily performed by numerical methods.

\section{Bipartite entanglement}
Let us consider two systems $A$ and $B$, represented by separable or inseparable Hilbert spaces $\mathcal{H}_A$ and $\mathcal{H}_B$.
The product space is $\mathcal{H}=\mathcal{H}_A\otimes\mathcal{H}_B$.
In our last contribution~\cite{vospe}, we introduced optimized conditions for entanglement in terms of arbitrary Hermitian operators.
A quantum state $\hat{\varrho}$ is entangled, iff there exists a bounded Hermitian operator $\hat{A}$ such that
\begin{align}
{\rm tr}(\hat{\varrho}\hat{A})>f_{AB}(\hat{A}),\label{hermcrit}
\end{align}
with $f_{AB}(\hat{A})=\sup\{{\rm tr}(\hat{\sigma}\hat{A}): \hat{\sigma} \ separable \}$ being the least upper bound of expectation values for separable states $\hat{\sigma}$.

Here, the set of positive integers is given as $\mathbb{N}=\{1,2,3,\dots\}$.
We define the countable set $S=\mathbb{N}\cup\{\infty\}$.
The cardinalities of the set of all natural numbers and $S$ are identical, $|S|=|\mathbb{N}|$.
Note the fact from set theory that the cardinalities of the sets $S$ and $S^2=S\times S$ are equal.
This means they are both countable.

An arbitrary mixed quantum state is represented by the statistical operator $\hat{\rho}$.
The spectral decomposition of the compact operator $\hat{\rho}$ reads as
\begin{align}
\hat{\rho}=\sum_{k=1}^N p_k |\psi_k\rangle\langle \psi_k|,\label{specdecomp}
\end{align} 
with $N\in S$, $(p_k)_{k=1,2,\dots,N}$ a finite or infinite positive sequence, which converges to $0$, and the orthonormalized vectors $|\psi_k\rangle$.
Each vector can be given in the Schmidt decomposition
\begin{align}
|\psi_k\rangle=\sum_{l=1}^{N_k} \lambda_{k,l} |a_{k,l}\rangle\otimes|b_{k,l}\rangle,\label{schmdecomp}
\end{align}
with $N_k\in S$, cf.~\cite{book1}.
With the following Theorem~\ref{lemma1} we can reduce the entanglement problem of even uncountable Hilbert spaces to separable Hilbert spaces.

\begin{theorem}\label{lemma1} 
For each state $\hat{\rho}$, exists a projection $\hat{P}$ on a separable subspace of $\mathcal{H}$ with the structure $V_A\otimes V_B$, which projects the state to itself: $\hat{P}\hat{\rho}\hat{P}=\hat{\rho}$.
\begin{description}
\item[Proof:] If the Hilbert space $\mathcal{H}$ is separable, then the conjecture is trivial.
Now let us assume, that $\mathcal{H}$ has an uncountable basis.
From Eqs.~(\ref{specdecomp})~and~(\ref{schmdecomp}) we conclude, that the range of $\hat{\rho}$ is a subspace of the Hilbert space $V_A\otimes V_B$, where $V_A$ and $V_B$ are generated by $|a_{k,l}\rangle$ and $|b_{k',l'}\rangle$ ($(k,l),(k',l')\in S^2$), respectively.
Due to the fact that $S\times S=S^2$ is countable,  $V_A$ and $V_B$ are separable Hilbert spaces.
It follows that the projection in the form $\hat{P}=\hat{P}_A\otimes\hat{P}_B$ to the space $V_A\otimes V_B$ satisfies $\hat{P}\hat{\rho}\hat{P}=\hat{\rho}$.
\end{description}
\end{theorem}

A conclusion of Theorem~\ref{lemma1} is, that we only have to consider separable Hilbert spaces $\mathcal{H}=\mathcal{H}_A\otimes\mathcal{H}_B$.
Thus the orthonormalized basis of $\mathcal{H}_A$ and $\mathcal{H}_B$ are $\{|e_k\rangle\}_{k\in\mathbb{N}}$ and $\{|f_l\rangle\}_{l\in\mathbb{N}}$, respectively.
Let us consider projections of $\mathcal{H}_A$ and $\mathcal{H}_B$ to finite subsystems $V_A$ and $V_B$. The operator corresponding to this local projection is $\hat{P}_s: \mathcal{H}_s \rightarrow V_s$ ($s=A,B$).
Now we can define the reduced quantum state:
\begin{align}
\hat{\rho}_{\rm red}=\hat{\rho}|_{V_A\otimes V_B}= (\hat{P}_A\otimes\hat{P}_B)\hat{\rho}(\hat{P}_A\otimes\hat{P}_B).\label{redstate}
\end{align}
The density operator is restricted to the finite space given as $V_A\otimes V_B$.
An example for such a projection could be for each $d\in\mathbb{N}$
\begin{align}
\hat{P}_A\otimes\hat{P}_B=\hat{P}_d=\left(\sum_{k=1}^{d}|e_k\rangle\langle e_k|\right)\otimes\left(\sum_{l=1}^{d}|f_l\rangle\langle f_l|\right).
\end{align}
In the following Lemma we construct a sequence of reduced operators converging to the statistical operator of the state.

\begin{lemma}\label{lemma2}
The sequence $\hat{\rho}_{{\rm red},d}=\hat{P}_d\hat{\rho}\hat{P}_d$, with the dimension $d\in\mathbb{N}$, converges for $d\to\infty$: $\hat{\rho}_{{\rm red},d} \to \hat{\rho}$.
\begin{description}
\item[Proof:]
According to the spectral decomposition of $\hat{\rho}$, Eq.~(\ref{specdecomp}), the state can be written as 
\begin{align*}
\hat{\rho}=\sum_{q=1}^\infty p_q |\psi_q\rangle\langle\psi_q|.
\end{align*}
The series $(\sum_{q=1}^r p_q)_{r\in\mathbb{N}}$ converges absolutely.
Hence we can write
\begin{align*}
\hat{\rho}_{{\rm red},d}&=\hat{P}_d\hat{\rho}\hat{P}_d\\
&=\sum_{q=1}^\infty p_q \hat{P}_d |\psi_q\rangle\langle\psi_q| \hat{P}_d.
\end{align*}
The sequence $(\hat{P}_d)_{d\in\mathbb{N}}$ converges pointwise to the identity operator $\hat{1}$.
This means for all $|\psi\rangle\in\mathcal{H}$: $\hat{P}_d|\psi\rangle\to|\psi\rangle$ for $d\to\infty$.
Therefore we obtain
\begin{align*}
\hat{\rho}_{{\rm red},d}=\sum_{q=1}^\infty p_q \hat{P}_d |\psi_q\rangle\langle\psi_q| \hat{P}_d \to \hat{\rho}.\label{projd}
\end{align*}
\end{description}
\end{lemma}

From Eq.~(\ref{hermcrit}) it follows that a state $\hat{\sigma}$ is separable, iff for all bounded Hermitian operators $\hat{A}$: ${\rm tr}(\hat{\sigma}\hat{A})\leq f_{AB}(\hat{A})$.
Due to Lemma~\ref{lemma2}, we have a sequence of compact operators with a finite dimensional range and this sequence converges to $\hat{\rho}$.
These facts enable us to prove, whether this sequence of reduced states reveals the entanglement of the state or not.

\begin{theorem}\label{theorem1}
For each entangled state $\hat{\varrho}$ there exist finite subspaces $V_A$ and $V_B$ of $\mathcal{H}_A$ and $\mathcal{H}_B$, respectively, so that the reduced state $\hat{\varrho}|_{V_A\otimes V_B}$ is entangled.
\begin{description}
\item[Proof:]
According to Theorem~\ref{lemma1} we can assume both systems $\mathcal{H}_A$ and $\mathcal{H}_B$ being separable.
Due to Lemma~\ref{lemma2} we can state, that $(\hat{\varrho}_{{\rm red},d})_{d\in\mathbb{N}}$ converges to $\hat{\varrho}$.
Now let us assume, that there exists an entangled state $\hat{\varrho}$, with $(\hat{\varrho}_{{\rm red},d})_{d\in\mathbb{N}}$ being a sequence of separable states.
This means ${\rm tr}(\hat{\varrho}_{{\rm red},d}\hat{A})\leq f_{AB}(\hat{A})$ for all bounded Hermitian operators $\hat{A}$.
For any bounded operator $\hat{A}$ the function $L(\hat{\rho})={\rm tr}(\hat{\rho}\hat{A})$ is continuous.
Thus we can conclude:
\begin{align*}
{\rm tr}(\hat{\varrho}\hat{A})&=L(\hat{\varrho})=L(\lim_{d\to\infty}\hat{\varrho}_{{\rm red},d})\\ &=\lim_{d\to\infty}\underbrace{L(\hat{\varrho}_{{\rm red},d})}_{\leq f_{AB}(\hat{A})}\leq f_{AB}(\hat{A}).
\end{align*}
This is equivalent to the separability of $\hat{\varrho}$ and in contradiction to the assumption, that $\hat{\varrho}$ is entangled.
Therefore there exists a positive integer $d_0$, with $\hat{\rho}_{{\rm red}, d_0}$ being entangled.
We can select $V_A$ and $V_B$ being generated by $\{|e_d\rangle\}_{d=1,2,\dots,d_0}$ and $\{|f_d\rangle\}_{d=1,2,\dots,d_0}$, respectively.
\end{description}
\end{theorem}

The existence of a reduced finite entangled state $\hat{\varrho}_{\rm red}$, identifies entanglement of $\hat{\varrho}$, because it exists an positive operator $\hat{C}$ (cf.~\cite{vospe}), with
\begin{align}
f_{AB}(\hat{C})<{\rm tr}(\hat{\varrho}_{\rm red}\hat{C})\leq{\rm tr}(\hat{\varrho}\hat{C}).
\end{align}
The opposite direction of this implication has been demonstrated in Theorem~\ref{theorem1}.
Now we can formulate the following corollary, which is the main result of our paper.

\begin{corollary}\label{corollary1}
A state $\hat{\varrho}$ is entangled, iff there exist finite subspaces $V_A$ and $V_B$ of $\mathcal{H}_A$ and $\mathcal{H}_B$, respectively, so that the reduced state $\hat{\varrho}|_{V_A\otimes V_B}$ is entangled.
\end{corollary}

Due to Corollary~\ref{corollary1} the entanglement problem is a problem in finite dimensional Hilbert spaces.
The dimension of the Hilbert space $\mathcal{H}_A\otimes\mathcal{H}_B$ is not important to formulate a solution.
Even for uncountable Hilbert spaces the problem can be solved in finite subspaces.
Explicitly, all $d_A\otimes d_B$-dimensional quantum systems, $d_A,d_B\in\mathbb{N}$, are necessary and sufficient to characterize bipartite entanglement.

May we consider to construct a counter example.
So let us assume, we have a sequence of entangled states 
\begin{align}
|\chi_k\rangle=\frac{1}{\sqrt{2}}(|0\rangle\otimes|0\rangle+|k\rangle\otimes|k\rangle),\quad (k\geq 1).
\end{align}
For each $k$, this is a Bell state and therefore entangled.
Therefore the state $|\chi_\infty\rangle=\lim_{k\to\infty} |\chi_k\rangle$ should be entangled.
We can conclude from the proof of Lemma~\ref{lemma2}, that the sequence of projections $\hat{P}_d$, see Eq.~(\ref{projd}), is necessary and sufficient to detect the entanglement.
But for any finite $d$ the projection is
\begin{align}
\hat{P}_d |\chi_\infty\rangle=\frac{1}{\sqrt{2}}|0\rangle\otimes|0\rangle,
\end{align}
which is a factorizeable vector.
This seem to be a contradiction to Theorem~\ref{theorem1}.
But $|\chi_\infty\rangle=\sum_{p,q=1}^\infty \chi_{p,q} |p\rangle\otimes|q\rangle$ should be an element of $\mathcal{H}$.
Therefore the state is normalized 
\begin{align}
\langle\chi_\infty|\chi_\infty\rangle=\sum_{p,q=1}^\infty |\chi_{p,q}|^2<\infty.
\end{align}
A necessary condition for the existence of $\sum_{p,q=1}^\infty |\chi_{p,q}|^2$ is, that $\chi_{p,q}\to 0$ for $p,q\to\infty$, which cannot be fullfiled for a possible state $\lim_{k\to\infty} |\chi_k\rangle$.
This state $|\chi_\infty\rangle$ simply does not exist, because the sequence $(|k\rangle)_{k\in\mathbb{N}}$ does not converge.
According to Corollary~\ref{corollary1}, every construction of such an example fails to be entangled or it fails to be a quantum state.
Entanglement is a problem, which is completely characterized in finite dimensional spaces.
Continuous variable systems do not need to be considered.
Furthermore, Corollary~\ref{corollary1} holds true even for continuous-variables bound entangled states.
With other words, the entanglement of any state is already revealed in a finite subsystem.

\section{multipartite entanglement}
Now we want to generalize the ideas, which has been performed so far.
Let us consider a compound system $\mathcal{H}$, consiting of all systems, which are elements of the set $\mathcal{M}$.
Each system is given by the Hilbert space $\mathcal{H}_s$ ($s\in\mathcal{M}$) and the compound system reads as the product $\mathcal{H}=\bigotimes_{s\in\mathcal{M}} \mathcal{H}_s$.

In this multipartite system a state $\hat{\sigma}$ is (fully) separable by definition, if it can be written as
\begin{align}
\hat{\sigma}=\sum_k p_k \bigotimes_{s\in\mathcal{M}} \hat{\rho}_{s,k},\label{defwerner}
\end{align}
with $p_k\geq 0$, $\sum_k p_k=1$ and the arbitrary quantum states $\hat{\rho}_{s,k}$ in the system $\mathcal{H}_s$, analogous to the definition by Werner~\cite{werner}.
A multipartite state is entangled, if it can not be written as Eq.~(\ref{defwerner}).
For example, the state
\begin{align}
|\psi\rangle=|0_A\rangle\otimes\frac{1}{\sqrt{2}}(|0_B\rangle\otimes|0_C\rangle+|1_B\rangle\otimes|1_C\rangle)\label{genex}
\end{align}
is (partially) entangled according to the definition given by Eq.~(\ref{defwerner}).
In this contribution a state is inseparable, if any partion is entangled.
In the example (\ref{genex}) the partion $BC$ is entangled.

Now let us consider a mixed state
\begin{align}
\hat{\rho}=\sum_{k=1}^N p_k |\psi_k\rangle\langle\psi_k|,
\end{align}
with $N\in S$.
From functional analysis and algebra we know that any vector $|\psi_k\rangle\in\mathcal{H}$ can be written as
\begin{align}
|\psi_k\rangle=\sum_{l=1}^{N_k} \bigotimes_{s\in\mathcal{M}} |a_{s,k,l}\rangle,
\end{align}
where the factorizeable vectors $\bigotimes_{s\in\mathcal{M}} |a_{s,k,l}\rangle$ are in general neither orthogonal nor normalized and $N_k\in S$.
So this is not a Schmidt decomposition, but a countable superposition of factorizeable vectors.
Again we obtain for each $s\in\mathcal{M}$ the space generated by $|a_{s,k,l}\rangle$ with $(k,l)\in S^2$.
Due to this we can assume all Hilbert spaces $\mathcal{H}_s$ being separable.
Therefore all theorems and lemmas hold true for multipartite entanglement.
And we can formulate the following Corollary.
\begin{corollary}\label{corollary2}
A multipartite state $\hat{\varrho}$ is entangled, iff there exist finite subspaces $V_s$ of $\mathcal{H}_s$, $s\in\mathcal{M}$, so that a reduced state $\hat{\varrho}|_{\bigotimes_{s\in\mathcal{M}} V_s}$ is entangled.
\end{corollary}

Note the fact, that the set $\mathcal{M}$ is not restricted to be countable.
For instance let us assume, that harmonic oscillators for any frequency $\omega>0$ are under study.
Each oscillator can be represented by the Fock basis of $\{|n_\omega\rangle\}_{n=0,1,\dots}$.
Due to $\omega\in\mathbb{R}$, we obtain the uncountable set
\begin{align}
\mathcal{M}=\{\omega\in\mathbb{R}: \omega>0 \}=]0, \infty[.
\end{align}
To verify entanglement, it is sufficient to identify the inseparability of a two-frequency partition.

For the proof of Theorem~\ref{theorem1} only one special sequence of $\hat{\varrho}_{{\rm red},d}$ is needed to identify the entanglement of the state.
For simplicity let us consider $\mathcal{M}=\{1,2,\dots,M\}$ ($M\in\mathbb{N}$) and the Hilbert spaces $\mathcal{H}_s$ ($s\in\mathcal{M}$) be separable.
An algorithm could be a sequence of $\hat{P}_d$ projections converging pointwise to $\hat{1}$.
Therefore the projection could be given in terms of the bases $\{|e_{s,l}\rangle\}_{l\in\mathbb{N}}$ as 
\begin{align}
\hat{P}_d=\bigotimes_{s\in\mathcal{M}} \hat{P}_{s,d}=\hat{P}_{1,d}\otimes\dots\otimes\hat{P}_{M,d},
\end{align}
with $\hat{P}_{s,d}=\sum_{l=1}^d|e_{s,l}\rangle\langle e_{s,l}|$.

Now we can check, if $\hat{\varrho}_{{\rm red},d}$ is entangled.
If it is not entangled, then increase the dimension $d$.
In this form Corollary~\ref{corollary2} reads as: The state is entangled, iff there exists a finite $d$, so that entanglement can be verified.
Then we obtained an integer $d$ and the algorithm stops.

\section{Conclusions}
In conclusion, we have given the proof that the general bipartite entanglement problem can be formulated in terms of finite Hilbert spaces.
Thus without any lost of generality we can assume finite spaces $d_A\otimes d_B$ to characterize entanglement.
Entanglement which requires infinite dimensions does not exist.
Furthermore we generalized these bipartite statement to multipartite entanglement.
We proposed an algorithm, which identifies multipartite entanglement for continuous variable systems.

\end{document}